\documentclass[a4paper,11pt]{article}

\usepackage{amsmath,amssymb}     
\usepackage{cite}                
\usepackage{hyperref}            

\numberwithin{equation}{section}   

\def \be {\begin{equation}}
\def \ee {\end{equation}}
\def \ba {\begin{array}}
\def \ea {\end{array}}
\def \bea{\begin{eqnarray}}
\def \eea{\end{eqnarray}}
\def \nn {\nonumber}

\def \a {\alpha}
\def \b {\beta}
\def \g {\gamma}

\def \d {\delta}

\def \e {\epsilon}

\def \m {\mu}
\def \n {\nu}

\def \s {\sigma}

\def \th {\theta}
\def \vth {\vartheta}

\def \t {\tau}
\def \z {\zeta}

\def \mA {\mathcal A}
\def \mB {\mathcal B}

\def \mN {\mathcal N}

\def \mP {\mathcal P}

\def \p {\partial}
\def \f {\frac}

\def \lt {\left}
\def \rt {\right}

\def \sr {\sqrt}

\def \ph  {\phantom}

\def \hi  {{\hat\imath}}
\def \hj  {{\hat\jmath}}

\def \dd {\mathrm{d}}

\def \ii {\mathrm{i}}

\def \Tr {{\textrm{Tr}}}

\def \diag {{\textrm{diag}}}

\def \and {{\textrm{and}}}

\def \GY {{\textrm{GY}}}
\def \DT {{\textrm{DT}}}

\begin{document}

\title{\textbf{Novel BPS Wilson loops in three-dimensional quiver Chern-Simons-matter theories}}
\author{
Hao Ouyang\footnote{ouyangh@ihep.ac.cn},
Jun-Bao Wu\footnote{wujb@ihep.ac.cn}~
and
Jia-ju Zhang\footnote{jjzhang@ihep.ac.cn}
}
\date{}

\maketitle

\vspace{-10mm}

\begin{center}
{\it
 Theoretical Physics Division, Institute of High Energy Physics, Chinese Academy of Sciences,\\
19B Yuquan Rd, Beijing 100049, China\\ \vspace{1mm}
Theoretical Physics Center for Science Facilities, Chinese Academy of Sciences,\\19B Yuquan Rd, Beijing 100049, China
}
\vspace{10mm}
\end{center}

\begin{abstract}

  We show that generic three-dimensional $\mathcal N=2$ quiver super Chern-Simons-matter theories admit Bogomol'nyi-Prasad-Sommerfield (BPS) Drukker-Trancanelli (DT) type Wilson loops. We investigate both Wilson loops along timelike infinite straight lines in Minkowski spacetime and circular Wilson loops in Euclidean space. In Aharnoy-Bergman-Jafferis-Maldacena theory, we find that generic BPS DT type Wilson loops preserve the same number of supersymmetries as Gaiotto-Yin type Wilson loops. There are several free parameters for generic BPS DT type Wilson loops in the construction, and supersymmetry enhancement for Wilson loops happens for special values of the parameters.

\end{abstract}

\baselineskip 18pt

\thispagestyle{empty}

\newpage

\tableofcontents

\section{Introduction}

Construction and classification of Bogomol'nyi-Prasad-Sommerfield (BPS) Wilson loops are certainly important subjects in the study of supersymmetric gauge theories.
The situation in three dimensions is more complicated than the four-dimensional case.
Gaiotto and Yin constructed BPS Wilson loops (along a straight line or a circle) in $\mN=2$ and $\mN=3$ super Chern-Simons-matter (CSM) theories by including scalar fields \cite{Gaiotto:2007qi}. This construction is quite similar to 1/2 BPS Wilson loops in four-dimensional $\mN=4$ super Yang-Mills theory \cite{Maldacena:1998im,Rey:1998ik}.
The idea of Gaiotto and Yin was adopted in \cite{Drukker:2008zx,Chen:2008bp,Rey:2008bh} to construct BPS Wilson loops in Aharnoy-Bergman-Jafferis-Maldacena (ABJM) theory \cite{Aharony:2008ug}. With some surprise, people only found 1/6 BPS Wilson loops within the class of Gaiotto-Yin (GY) type.
However, the study of dual fundamental string solutions in AdS$_4\times$CP$^3$ \cite{Drukker:2008zx,Rey:2008bh} indicates that there should be 1/2 BPS Wilson loops in ABJM theory.
About one year later, such Wilson loops were finally constructed by Drukker and Trancanelli \cite{Drukker:2009hy} via including fermions in a clever way. This construction was explained elegantly through the Brout-Englert-Higgs (BEH) mechanism in \cite{Lee:2010hk}, and 2/5 BPS Drukker-Trancanelli (DT) type Wilson loops in $\mN=5$ CSM theories \cite{Hosomichi:2008jb,Aharony:2008gk} were also constructed in this paper. Later DT type 1/2 BPS Wilson loops in $\mN=4$ CSM theories were constructed in \cite{Ouyang:2015qma,Cooke:2015ila}. On the other hand, GY type Wilson loops generically preserve two Poincar\'e supercharges. The only known exceptional case is the Wilson loops associated with certain end of $\mN=4$ linear quiver theories, where supersymmetry (SUSY) enhancement appears  \cite{Cooke:2015ila}.

The previous results may tend to let people assume that DT type Wilson loops are very rare and their existence requires that the theory have a quite large number of supersymmetries. DT type Wilson loops also seem to preserve more supersymmetries than the GY type Wilson loops when they are along the same curve.
The result in this letter will show that it is not the case. For generic $\mN=2$ quiver CSM theories, for each bifundamental (or adjoint) matter chiral multiplet we can construct 1/2 BPS DT type Wilson loops.
The BPS Wilson loops include the ones along timelike straight lines in Minkowski spacetime and the circular ones in Euclidean space.
This construction can be generalized to CSM theories with more supersymmetries. In ABJM theory, we find that generic DT type BPS Wilson loops are 1/6 BPS, and they preserve the same number of supersymmetries as GY type Wilson loops.
There are several free parameters in the generic 1/6 BPS DT type Wilson loops, and for special values of the parameters the preserved supersymmetries are enhanced to 1/2 BPS. The generic Wilson loops constructed here are not invariant under local $SU(3)$ transformations which form a subgroup of the $SU(4)$ R-symmetry. This is a big difference from the previously constructed less BPS Wilson loops in ABJM theory \cite{Griguolo:2012iq,Cardinali:2012ru,Kim:2013oza,Bianchi:2014laa,Correa:2014aga}. In this short letter, we only give the main results and the details of the derivations and some generalizations are presented in \cite{Ouyang:2015bmy}.

\section{$\bf{\mN=2}$ quiver CSM theories}

We consider generic $\mN=2$ quiver CSM theories with bifundamental matters. We pick two adjacent nodes in the quiver diagram and assume that the corresponding gauge groups are $U(N)$ and $U(M)$. 
The vector multiplet for gauge group $U(N)$ includes $A_\mu, \sigma, \chi, D$ with the last three ones being auxiliary fields. Similarly, for the gauge group $U(M)$, we have the vector multiplet with fields $\hat{A}_\mu, \hat{\chi}, \hat{\sigma}, \hat{D}$.
The chiral multiplet in the bifundamental representation of $U(N)\times U(M)$ consists of the scalar $\phi$, spinor $\psi$ and auxiliary field $F$. There could be additional matters couple to these two gauge fields. They will enter into  the on-shell values of $\sigma$ and $ \hat{\sigma}$ in the Wilson loops which we will construct. However the structure of these Wilson loops will  not be affected.

The SUSY transformations of $\mN=2$ CSM theory could be found in \cite{Schwarz:2004yj}. For the vector multiplet part, we only need the following off-shell SUSY transformations of $A_\mu, \sigma, \hat{A}_\mu, \hat{\sigma}$,
\bea \label{offshell}
&&\hspace{-5mm} \d A_\m=\f{1}{2} (\bar\chi \g_\m\e+\bar\e \g_\m\chi), ~~~
                \d \s=-\f{\ii}{2} (\bar\chi\e+\bar\e\chi),                                     \nn\\
&&\hspace{-5mm} \d \hat A_\m=\f{1}{2} (\bar{\hat\chi} \g_\m\e+\bar\e \g_\m\hat\chi), ~~~
                \d \hat\s=-\f{\ii}{2} (\bar{\hat\chi}\e+\bar\e\hat\chi),
\eea
and for the matters part we only need the off-shell SUSY transformations of $\phi$ and $\psi$,
\bea
&& \d \phi = \ii\bar\e\psi, ~~~ \d\bar\phi=\ii\bar\psi\e,  \nn\\
&& \d\psi = (-\g^\m D_\m \phi -\s \phi+\phi\hat\s)\e -\vth\phi+ \ii \bar\e F ,  \\
&& \d\bar\psi = \bar\e( \g^\m D_\m \bar\phi +\hat\s\bar\phi -\bar\phi\s)-\bar\vth\bar\phi-\ii \e\bar F, \nn
\eea
with the definitions of covariant derivatives
\bea
&& D_\m \phi=\p_\m\phi+\ii A_\m\phi-\ii\phi \hat A_\m,        \nn\\
&& D_\m \bar\phi=\p_\m\bar\phi+\ii \hat A_\m\bar\phi-\ii\bar\phi A_\m.
\eea
The SUSY parameters are $\e=\th+x^\m\g_\m\vth$, $\bar\e=\bar\th-\bar\vth x^\m\g_\m$, with $\th$, $\bar\th$ denoting Poincar\'e SUSY and $\vth$, $\bar\vth$ denoting superconformal symmetry.

As shown in \cite{Ouyang:2015ada}, BPS Wilson loops in Minkowski spacetime cannot be spacelike. In this letter, we consider Wilson loops along timelike infinite straight lines in Minkowski spacetime.
For a timelike infinite straight line in Minkowski spacetime, Wick rotation makes it an infinite straight line in Euclidean space, and then a suitable conformal transformation can change it into a circle.
So in a superconformal theory, BPS Wilson loops along timelike infinite straight lines in Minkowski indicate the existence of circular BPS Wilson loops in Euclidean space. We will also construct such circular Wilson loops in this letter.

In Minkowski spacetime, one can construct a GY type 1/2 BPS Wilson loop along the timelike infinite straight line $x^\m=\t\d^\m_0$ \cite{Gaiotto:2007qi}
\bea \label{gy1}
&& W_{\GY}=\mP \exp \lt( -\ii\int\dd\t L_{\GY}(\t) \rt),\\
&& L_\GY=\lt( \ba{cc} A_\m\dot x^\m+\s |\dot x| & \\ & \hat A_\m\dot x^\m+\hat\s |\dot x| \ea \rt).\nn
\eea
For BPS Wilson loops along straight lines, Poincar\'e SUSY and conformal SUSY are separately preserved and similar, and so it is enough to consider only Poincar\'e SUSY.
The preserved Poincar\'e supersymmetries are given by
\be \label{susy1}
\g_0\th=\ii\th, ~~~ \bar\th\g_0=\ii\bar\th.
\ee

We now construct the DT type Wilson loop along $x^\m=\t \d^\m_0$
\bea
&& W_{\DT}=\mP \exp \lt( -\ii\int\dd\t L_{\DT}(\t) \rt),                        ~~~
   L_\DT=\lt( \ba{cc} \mA & \bar f_1 \\ f_2 & \hat\mA \ea \rt),                 \nn\\
&& \mA=A_\m\dot x^\m+\s |\dot x|+ m \phi\bar\phi |\dot x|, ~~~
   \bar f_1=\bar\zeta\psi |\dot x|,                                             \\
&& \hat\mA=\hat A_\m\dot x^\m+\hat\s |\dot x| + n \bar\phi\phi |\dot x|, ~~~
   f_2=\bar\psi\eta |\dot x|.                                                   \nn
\eea
To make it preserve the supersymmetries (\ref{susy1}), it is enough to require that \cite{Lee:2010hk}
\be
\d L_\DT=\p_\t G+\ii[L_\DT,G], \label{lee0}
\ee
for some Grassmann odd matrix
\be
G= \lt( \ba{cc} & \bar g_1 \\ g_2 &  \ea \rt).
\ee
Concretely, one needs
\bea \label{lee}
&& \d \mA=\ii(\bar f_1 g_2-\bar g_1 f_2),   \nn\\
&& \d \hat\mA=\ii(f_2 \bar g_1 - g_2 \bar f_1),\\
&& \d \bar f_1 = \p_\t \bar g_1+\ii\mA \bar g_1-\ii\bar g_1\hat\mA,\nn\\
&& \d f_2 = \p_\t g_2+\ii\hat\mA g_2-\ii g_2\mA. \nn
\eea
We find  that the necessary and sufficient conditions for the existence of such $\bar g_1$ and $g_2$ are
\be
\bar\zeta^\a=\bar\a (1,\ii), ~~~
\eta_\a=(1,-\ii)\b, ~~~
m=n=2\ii\bar\a\b.
\ee
Such a DT type Wilson loop is 1/2 BPS, and the preserved supersymmetries are (\ref{susy1}). Note that there are two free complex parameters $\bar\a$ and $\b$ , and they can be any complex constants. When $\bar\a=\b=0$, it goes back to the GY type Wilson loop.

Similarly, in Euclidean space we have the 1/2 BPS GY type and DT type Wilson loops along the circle $x^\m=(\cos\t,\sin\t,0)$.
The GY type Wilson loop is
\bea \label{z1}
&& W_{\GY}=\Tr\mP \exp \lt( -\ii\oint\dd\t L_{\GY}(\t) \rt),\\
&& L_\GY=\lt( \ba{cc} A_\m\dot x^\m-\ii\s |\dot x| & \\ & \hat A_\m\dot x^\m-\ii\hat\s |\dot x| \ea \rt),\nn
\eea
and the DT type Wilson loop is
\bea
&& W_{\DT}=\Tr\mP \exp \lt( -\ii\oint\dd\t L_{\DT}(\t) \rt),                        ~~~
   L_\DT=\lt( \ba{cc} \mA & \bar f_1 \\ f_2 & \hat\mA \ea \rt),                 \nn\\
&& \mA=A_\m\dot x^\m -\ii\s |\dot x| -2\bar\a\b\phi\bar\phi |\dot x|, ~~~
   \bar f_1=\bar\zeta\psi |\dot x|,                                             \nn\\
&& \hat\mA=\hat A_\m\dot x^\m -\ii\hat\s |\dot x| -2\bar\a\b \bar\phi\phi |\dot x|, ~~~
   f_2=\bar\psi\eta |\dot x|.                                                   \\
&& \bar\z^\a=\bar\a(e^{\ii\t/2},e^{-\ii\t/2}), ~~~
   \eta_\a=(e^{-\ii\t/2},e^{\ii\t/2})\b.                                        \nn
\eea
The 1/2 BPS GY and DT type Wilson loops preserve the same supersymmetries
\be \label{z2}
\vth=\ii \g_3\th, ~~~ \bar\vth=\bar\th\ii\g_3.
\ee

We would like to point out that this construction can also be applied to the case when $U(N)\times U(M)$ is replaced by $SO(N)\times USp(2M)$, and the case when there are matter fields in adjoint representation. For the latter case, one just simply sets $\hat{A}_\mu \equiv A_\mu$ and $\hat{\sigma}\equiv\sigma$.

\section{$\mN=2$ quiver CSM theories with multiple matters}

Now we turn to the case when there are multiple matter fields in the bifundamental and anti-bifundamental  representations in the $\mN=2$ theory.
These multiplets include fields $\phi_i, \psi_i, F_i$ and $\phi_\hi, \psi_\hi, F_\hi$, respectively.
The off-shell SUSY transformations of the gauge fields part (\ref{offshell}) do not change, and those of the matters part include
\bea
&& \d \phi_i = \ii\bar\e\psi_i, ~~~ \d\bar\phi^i=\ii\bar\psi^i\e, \nn\\
&& \d\psi_i = (-\g^\m D_\m \phi_i -\s \phi_i+\phi_i\hat\s)\e -\vth\phi_i + \ii \bar\e F_i , \nn\\
&& \d\bar\psi^i = \bar\e( \g^\m D_\m \bar\phi^i +\hat\s\bar\phi^i -\bar\phi^i\s) -\bar\vth\bar\phi^i-\ii \e\bar F^i,\nn\\
&& \d \phi_\hi = \ii\bar\e\psi_\hi, ~~~ \d\bar\phi^\hi=\ii\bar\psi^\hi\e,\\
&& \d\psi_\hi = (-\g^\m D_\m \phi_\hi -\hat\s \phi_\hi+\phi_\hi\s)\e -\vth\phi_\hi+ \ii \bar\e F_\hi ,\nn\\
&& \d\bar\psi^\hi = \bar\e( \g^\m D_\m \bar\phi^\hi +\s\bar\phi^\hi -\bar\phi^\hi\hat\s) -\bar\vth\bar\phi^\hi -\ii \e\bar F^\hi. \nn
\eea
Here we have $i=1,2,\cdots,N_f$, and $\hi=\hat 1,\hat 2,\cdots,\hat N_{\hat f}$.
The definitions of the covariant derivatives are
\bea
&& D_\m \phi_i=\p_\m\phi_i+\ii A_\m\phi_i-\ii\phi_i \hat A_\m,                           ~~~
   D_\m \bar\phi_i=\p_\m\bar\phi_i+\ii \hat A_\m\bar\phi_i-\ii\bar\phi_i A_\m,           \nn\\
&& D_\m\phi_\hi=\p_\m\phi_\hi+\ii \hat A_\m\phi_\hi-\ii\phi_\hi A_\m,                    ~~~
   D_\m \bar\phi_\hi=\p_\m\bar\phi_\hi+\ii A_\m\bar\phi_\hi-\ii\bar\phi_\hi \hat A_\m.
\eea

In Minkowski spacetime we still have the GY type 1/2 BPS Wilson loop (\ref{gy1}), and the supersymmetries it preserves are (\ref{susy1}).
We construct the DT type Wilson loop along $x^\m=\t\d^\m_0$
\bea
&& W_{\DT}=\mP \exp \lt( -\ii\int\dd\t L_{\DT}(\t) \rt),                        ~~~
   L_\DT=\lt( \ba{cc} \mA & \bar f_1 \\ f_2 & \hat\mA \ea \rt),                 \nn\\
&& \mA=A_\m\dot x^\m+\s |\dot x|+ \mB|\dot x|,                                  ~~~
   \hat\mA=\hat A_\m\dot x^\m+\hat\s |\dot x| + \hat \mB|\dot x|,               \nn\\
&& \mB= M^i_{\ph i j}\phi_i\bar\phi^j+M_\hi^{\ph\hi\hj} \bar\phi^\hi\phi_\hj
                                   +M^{ i\hi}\phi_i\phi_\hi  + M_{\hi i}\bar\phi^\hi\bar\phi^i, \\
&&\hat \mB=  N_i^{\ph ij}\bar\phi^i\phi_j + N^\hi_{\ph\hi\hj}\phi_\hi\bar\phi^\hj
                                                 + N_{i\hi}\bar\phi^i\bar\phi^\hi+N^{\hi i}\phi_\hi\phi_i, \nn\\
&& \bar f_1=(\bar\zeta^i\psi_i + \bar\psi^\hi \m_\hi ) |\dot x|, ~~~
   f_2= (\bar\psi^i\eta_i+\bar\n^\hi\psi_\hi )|\dot x| .        \nn
\eea
We want it to preserve the supersymmetries (\ref{susy1}).
Starting from (\ref{lee}), we have the parameterizations
\bea
&& \bar\zeta^i=\bar\a^i\bar\zeta, ~~~
   \m_\hi=\m\g_\hi, ~~~
   \eta_i=\eta \b_i, ~~~
   \bar\n^\hi=\bar\d^\hi\bar\n,  \nn\\
&& \bar\zeta^\a = \bar\n^\a=(1,\ii), ~~~ \eta_\a=\m_\a=(1,-\ii),
\eea
with $\bar\a^i$, $\g_\hi$, $\b_i$, and $\d^\hi$ being complex constants, and we also have the following conditions
\bea\label{e1}
&& M^i_{\ph ij}=N_j^{\ph ji}=2\ii\bar\a^i\b_j, ~~~ M^{i\hi}=N^{\hi i}=\bar\a^i\bar\d^\hi=0,  \nn\\
&& M_\hi^{\ph\hi\hj}=N_\hj^{\ph \hj\hi}=2\ii\g_\hi\bar\d^\hj, ~~~ M_{\hi i}=N_{i\hi}=\g_\hi\b_i=0.
\eea
These lead to four classes of solutions.
\begin{itemize}

\item{Class I}
\be \label{gj1}
\g_\hi=\bar\d^\hi=0.
\ee

\item{Class II}
\be \label{gj2}
\bar\a^i=\b_i=0.
\ee

\item{Class III}
\be \label{gj3}
\b_i=\bar\d^\hi=0.
\ee

\item{Class IV}
\be \label{gj4}
\bar\a^i=\g_\hi=0.
\ee

\end{itemize}

In Euclidean space we still have the GY type 1/2 BPS Wilson loop (\ref{z1}), and the preserved supersymmetries are (\ref{z2}).
We construct the DT type Wilson loop along $x^\m=(\cos\t,\sin\t,0)$
\bea
&& W_{\DT}=\Tr\mP \exp \lt( -\ii\oint\dd\t L_{\DT}(\t) \rt),                        ~~~
   L_\DT=\lt( \ba{cc} \mA & \bar f_1 \\ f_2 & \hat\mA \ea \rt),                 \nn\\
&& \mA=A_\m\dot x^\m -\ii\s |\dot x|-2( \bar\a^i\b_j \phi_i\bar\phi^j + \g_\hi\bar\d^\hj \bar\phi^\hi\phi_\hj )|\dot x|,  \nn\\
&& \hat\mA=\hat A_\m\dot x^\m -\ii\hat\s |\dot x| - 2( \bar\a^i\b_j \bar\phi^j\phi_i + \g_\hi\bar\d^\hj \phi_\hj\bar\phi^\hi ) |\dot x|,\\
&& \bar f_1=(\bar\a^i\bar\zeta\psi_i + \bar\psi^\hi \eta\g_\hi ) |\dot x|, ~~~
   f_2= (\bar\psi^i\eta\b_i+\bar\d^\hi\bar\z\psi_\hi )|\dot x| ,                                                   \nn\\
&& \bar\zeta^\a = (e^{\ii\t/2},e^{-\ii\t/2}), ~~~ \eta_\a=(e^{-\ii\t/2},e^{\ii\t/2}).       \nn                                                \nn
\eea
Similar to the case in Minkowski spacetime, we have four classes of solutions (\ref{gj1}), (\ref{gj2}), (\ref{gj3}), and (\ref{gj4}) that make this circular DT type Wilson loop 1/2 BPS and preserve supersymmetries (\ref{z2}).

\section{ABJM theory}

The on-shell SUSY transformations of ABJM theory are \cite{Gaiotto:2008cg,Hosomichi:2008jb,Terashima:2008sy,Bandres:2008ry}
\bea \label{susytransfabjm}
&& \d A_\m=\f{4\pi}{k} \lt( \phi_I\bar\psi_J\g_\m\e^{IJ} +\bar\e_{IJ}\g_\m\psi^J\bar\phi^I \rt), ~~~
   \d\hat A_\m=\f{4\pi}{k} \lt( \bar\psi_J\g_\m\phi_I\e^{IJ}+\bar\e_{IJ}\bar\phi^I\g_\m\psi^J \rt), \nn\\
&& \d\phi_I=2i\bar\e_{IJ}\psi^J, ~~~ \d\bar\phi^I=2i\bar\psi_J\e^{IJ},\\
&& \d\psi^I=2\g^\m\e^{IJ}D_\m\phi_J +2\vth^{IJ}\phi_J
            -\f{4\pi}{k}\e^{IJ} \lt( \phi_J\bar\phi^K\phi_K-\phi_K\bar\phi^K\phi_J \rt)
            -\f{8\pi}{k}\e^{KL}\phi_K\bar\phi^I\phi_L, \nn\\
&& \d\bar\psi_I=-2\bar\e_{IJ}\g^\m D_\m\bar\phi^J +2\bar\vth_{IJ}\bar\phi^J
                +\f{4\pi}{k}\bar\e_{IJ} \lt( \bar\phi^J\phi_K\bar\phi^K-\bar\phi^K\phi_K\bar\phi^J \rt)
                +\f{8\pi}{k}\bar\e_{KL}\bar\phi^K\phi_I\bar\phi^L, \nn
\eea
with the SUSY parameters $\e^{IJ}=\th^{IJ}+x^\m\g_\m\vth^{IJ}$, $\bar\e_{IJ}=\bar\th_{IJ}-\bar\vth_{IJ}x^\m\g_\m$.
In Minkowski spacetime $\th^{IJ}$, $\bar\th_{IJ}$, $\vth^{IJ}$, $\bar\vth_{IJ}$ are Dirac spinors with constraints
\bea
&& \th^{IJ}=-\th^{JI}, ~~~ (\th^{IJ})^*=\bar \th_{IJ}, ~~~ \bar\th_{IJ}=\f{1}{2}\e_{IJKL}\th^{KL}, \nn\\
&& \vth^{IJ}=-\vth^{JI}, ~~~ (\vth^{IJ})^*=\bar \vth_{IJ}, ~~~ \bar\vth_{IJ}=\f{1}{2}\e_{IJKL}\vth^{KL}.
\eea
Symbol $\e_{IJKL}$ is totally antisymmetric with $\e_{1234}=1$. In Euclidean space the constraints become
\be
\th^{IJ}=-\th^{JI}, ~~~ \bar\th_{IJ}=\f{1}{2}\e_{IJKL}\th^{KL},~~~
\vth^{IJ}=-\vth^{JI},  ~~~ \bar\vth_{IJ}=\f{1}{2}\e_{IJKL}\vth^{KL}.
\ee
We have definitions of covariant derivatives
\be
D_\m\phi_J =\p_\m \phi_J +\ii A_\m \phi_J -\ii \phi_J \hat A_\m ,   ~~~
D_\m\bar\phi^J=\p_\m\bar\phi^J +\ii \hat A_\m \bar\phi^J -\ii \bar\phi^J  A_\m.
\ee

In Minkowski spacetime, a general GY type Wilson loop along $x^\m=\t\d^\m_0$ takes the form
\bea\label{abjmgy}
&& W_\GY=\mP \exp \lt( -\ii\int\dd\t L_\GY(\t) \rt), ~~~
   L_\GY=\lt( \ba{cc} \mA_\GY & \\ & \hat\mA_\GY \ea \rt),\\
&& \mA_\GY=A_\m \dot x^\m +\f{2\pi}{k} R^I_{\ph{I}J} \phi_I\bar\phi^J |\dot x|,           ~~~
   \hat\mA_\GY=\hat A_\m \dot x^\m +\f{2\pi}{k} S_I^{\ph{I}J} \bar\phi^I\phi_J |\dot x|.\nn
\eea
We find that,
up to some $SU(4)$ R-symmetry transformation, the only BPS GY type Wilson loop is the one with $R^I_{\ph IJ}=S_J^{\ph JI}=\diag(-1,-1,1,1)$. It is 1/6 BPS and preserves the supersymmetries
\bea\label{susy2}
&& \g_0\th^{12}=\ii\th^{12}, ~~~ \g_0\th^{34}=-\ii\th^{34},  \nn\\
&& \th^{13}=\th^{14}=\th^{23}=\th^{24}=0.
\eea
This is just the Wilson loop that was constructed in \cite{Drukker:2008zx,Chen:2008bp,Rey:2008bh}.  Especially, we find that we do not need to require that $R^I_{\ph I J}$ or $S_I^{\ph I J}$ is a hermitian matrix \emph{a priori}, and we can show that it is the result of SUSY invariance.

We turn to constructing a DT type Wilson loop that preserves at least the supersymmetries (\ref{susy2}). In Minkowski spacetime, a general DT type Wilson loop is \cite{Drukker:2009hy}
\bea
&& W_\DT = \mP \exp \lt( -\ii\int\dd\t L_\DT(\t) \rt), ~~~
   L_\DT = \lt( \ba{cc} \mA & \bar f_1 \\ f_2 & \hat\mA \ea \rt),                                   \nn\\
&& \mA = \mA_\GY +\f{2\pi}{k} M^I_{\ph{I}J} \phi_I\bar\phi^J |\dot x|, ~~~
   \bar f_1=\sr{\f{2\pi}{k}}\bar\zeta_I\psi^I |\dot x|,                                      \\
&& \hat\mA = \hat\mA_\GY +\f{2\pi}{k} N_I^{\ph{I}J} \bar\phi^I\phi_J |\dot x|, ~~~
   f_2=\sr{\f{2\pi}{k}}\bar\psi_I\eta^I |\dot x|.                                   \nn
\eea
The existence of $G$ in (\ref{lee0}) leads to the  parameterizations
\bea
&& \bar\z_{1,2}=\bar\a_{1,2}\bar\z, ~~~ \bar \z^\a=(1,\ii),    ~~~
   \bar\z_{3,4}=\bar\g_{3,4}\bar\m, ~~~ \bar \m^\a=(-\ii,-1),   \nn\\
&& \eta^{1,2}=\eta\b^{1,2}, ~~~ \eta_\a=(1,-\ii),              ~~~
   \eta^{3,4}=\n\d^{3,4}, ~~~ \n_\a=(-\ii,1),
\eea
and the following conditions
\bea
&& M^I_{\ph IJ}\bar\z_L\gamma_0\th^{LK}= M^K_{\ph KJ} \bar\z_L\gamma_0\theta^{LI},      ~~~
   M^I_{\ph IJ} \bar\th_{KL}\gamma_0\eta^L = M^I_{\ph IK} \bar\th_{JL}\gamma_0\eta^L,   \\
&& M^I_{\ph IK} \th^{KJ} = - \eta^J \bar\z_K\g_0\th^{KI},                             ~~~
   M^K_{\ph KI} \bar\th_{KJ} = -\bar\th_{KI}\g_0\eta^K \bar\zeta_J,                   ~~~
   M^I_{\ph IJ}=N_J^{\ph JI}.                                                         \nn
\eea
We find four classes of solutions, and all of them satisfy
\be
M^I_{\ph IJ}=2\ii\lt(\ba{cccc}
\bar\a_2\b^2  & -\bar\a_2\b^1 & &\\
-\bar\a_1\b^2 & \bar\a_1\b^1  & & \\
  &  & \bar\g_4\d^4  & -\bar\g_4\d^3 \\
  &  & -\bar\g_3\d^4 & \bar\g_3\d^3
\ea\rt).
\ee
\begin{itemize}
\item{Class I}
\be \label{jj1}
\bar\g_{3,4}=\d^{3,4}=0.
\ee

\item{Class II}
\be \label{jj2}
\bar\a_{1,2}=\b^{1,2}=0.
\ee

\item{Class III}
\be \label{jj3}
\b^{1,2}=\d^{3,4}=0.
\ee

\item{Class IV}
\be \label{jj4}
\bar\a_{1,2}=\bar\g_{3,4}=0.
\ee

\end{itemize}

These solutions generically lead to 1/6 BPS Wilson loops.
Within each class, we search for Wilson loops that preserve more supersymmetries.
The result is that such loops only appear in class I and class II.
In class I, we define
\be
\bar\a_I=(\bar\a_1,\bar\a_2,0,0), ~~~
   \b^I=(\b^1,\b^2,0,0).
\ee
The SUSY enhancement happens only when
\be \label{e4}
\b^I=-\f{\ii}{\bar\a_J \a^J} \a^I, ~~~ \bar\a_I\neq0,
\ee
with $\a^I\equiv\bar\a_I^*$.
Under this condition, the Wilson loop becomes 1/2 BPS and the preserved supersymmetries are
\be \label{s1}
\g_0 \bar\a_I \th^{IJ}=\ii\bar\a_I\th^{IJ},               ~~~
\g_0 \e_{IJKL}\a^J \th^{KL}=-\ii\e_{IJKL}\a^J \th^{KL}.
\ee
This kind of 1/2 BPS Wilson loops is essentially the ones that were constructed in \cite{Drukker:2009hy}, up to some possible $SU(4)$ R-symmetry transformations.

Similarly, in class II we define
\be
\bar\g_I=(0,0,\bar\g_3,\bar\g_4),  ~~~
\d^I=(0,0,\d^3,\d^4).
\ee
When \be
\d^I=\f{\ii}{\bar\g_J \g^J} \g^I, ~~~ \bar\g_I\neq0,
\ee
with $\g^I\equiv\bar\g_I^*$, the Wilson loop is 1/2 BPS, and the preserved supersymmetries are
\be
\g_0 \bar\g_I \th^{IJ}=-\ii\bar\g_I\th^{IJ},            ~~~
\g_0 \e_{IJKL}\g^J \th^{KL}=\ii\e_{IJKL}\g^J \th^{KL}.
\ee

Note that there are no Wilson loops  which are exactly 1/3 BPS within the class of Wilson loops that preserve supersymmetries including ones  given by (\ref{susy2}).

In Euclidean space, one can construct the circular 1/6 BPS GY type Wilson loop along $x^\m=(\cos\t,\sin\t,0)$
\bea
&& W_\GY=\Tr \mP\exp \lt( -\ii\oint\dd\t L_\GY(\t) \rt), ~~~
   L_\GY=\lt( \ba{cc} \mA_\GY & \\ & \hat\mA_\GY \ea \rt),                                 \nn\\
&& \mA_\GY=A_\m \dot x^\m +\f{2\pi}{k} R^I_{\ph{I}J} \phi_I\bar\phi^J |\dot x|,           ~~~
   \hat\mA_\GY=\hat A_\m \dot x^\m +\f{2\pi}{k} R^I_{\ph{I}J}\bar\phi^J\phi_I |\dot x|,    \\
&& R^I_{\ph{I}J}=\diag(\ii,\ii,-\ii,-\ii).                                                 \nn
\eea
The preserved supersymmetries are
\be \label{e22}
\vth^{12}=\ii\g_3\th^{12}, ~~~ \vth^{34}=-\ii\g_3\th^{34}.
\ee
We also construct the DT type Wilson loop along $x^\m=(\cos\t,\sin\t,0)$
\bea
&& W_\DT =\Tr \mP \exp \lt( -\ii\oint\dd\t L_\DT(\t) \rt), ~~~
   L_\DT = \lt( \ba{cc} \mA & \bar f_1 \\ f_2 & \hat\mA \ea \rt),                         \nn\\
&& \mA = A_\m\dot x^\m +\f{2\pi}{k} U^I_{\ph{I}J} \phi_I\bar\phi^J |\dot x|, ~~~
   \bar f_1=\sr{\f{2\pi}{k}}( \bar\a_I\bar\zeta + \bar\g_I\bar\m )\psi^I |\dot x|,         \nn\\
&& \hat\mA = \hat A_\m\dot x^\m +\f{2\pi}{k} U^I_{\ph{I}J}\bar\phi^J\phi_I |\dot x|, ~~~
   f_2=\sr{\f{2\pi}{k}}\bar\psi_I(\eta\b^I+\n\d^I) |\dot x|,                               \\
&& U^I_{\ph IJ}=\lt(\ba{cccc}
\ii-2\bar\a_2\b^2  & 2\bar\a_2\b^1      & &\\
2\bar\a_1\b^2      & \ii-2\bar\a_1\b^1  & & \\
                   &                    & -\ii-2\bar\g_4\d^4  & 2\bar\g_4\d^3 \\
                   &                    & 2\bar\g_3\d^4       & -\ii-2\bar\g_3\d^3
\ea\rt),                                                                                    \nn\\
&& \bar\a_I=(\bar\a_1,\bar\a_2,0,0), ~~~
   \bar\z^\a=(e^{\ii\t/2},e^{-\ii\t/2}), ~~~
   \b^I=(\b^1,\b^2,0,0), ~~~
   \eta_\a=(e^{-\ii\t/2},e^{\ii\t/2}),                                                   \nn\\
&& \bar\g_I=(0,0,\bar\g_3,\bar\g_4), ~~~
   \bar\m^\a=(e^{\ii\t/2},-e^{-\ii\t/2}), ~~~
   \d^I=(0,0,\d^3,\d^4), ~~~
   \n_\a=(-e^{-\ii\t/2},e^{\ii\t/2}). \nn
\eea
Similar to the case in Minkowski spacetime, we have four classes of solutions (\ref{jj1}), (\ref{jj2}), (\ref{jj3}), and (\ref{jj4}) that make this DT type Wilson 1/6 BPS, and the preserved supersymmetries are (\ref{e22}).
In class I, when
\be
\b^I=\f{\ii}{\bar\a_J \a^J} \a^I, ~~~ \bar\a_I\neq0,
\ee
the Wilson loop becomes 1/2 BPS, and the preserved supersymmetries are
\be
\bar\a_I \vth^{IJ}=\ii\g_3 \bar\a_I\th^{IJ},               ~~~
\e_{IJKL}\a^J \vth^{KL}=-\ii\g_3\e_{IJKL}\a^J \th^{KL}.
\ee
In class II when \be
\d^I=-\f{\ii}{\bar\g_J \g^J} \g^I, ~~~ \bar\g_I\neq0,
\ee
the Wilson loop is 1/2 BPS, and the preserved supersymmetries are
\be
\bar\g_I \vth^{IJ}=-\ii\g_3 \bar\g_I\th^{IJ},            ~~~
\e_{IJKL}\g^J \vth^{KL}=\ii\g_3 \e_{IJKL}\g^J \th^{KL}.
\ee

\section{Conclusion and Discussion}

The results in this letter show generic existence of DT type BPS Wilson loops in $\mN=2$ quiver CSM theories. We also find that generic DT type Wilson loops in ABJM theory  preserve the same number of supersymmetries as GY type Wilson loops.
There are free parameters in the DT type BPS Wilson loops, and only for special values of the parameters the Wilson loops preserve half of the supersymmetries, three times of the ones preserved by the GY type Wilson loops.
The new generic DT type BPS Wilson loops constructed here  are interpolating between GY type BPS Wilson loops and 1/2 BPS DT type Wilson loops.

The results in this letter can be generalized to $\mN=4$ and $\mN=3$ CSM theories.
For Wilson loops in $\mN=4$ CSM theories, generically both GY type and DT type BPS Wilson loops are 1/4 BPS, and for special cases the latter ones  preserve half of the supersymmetries.
Notice that for BPS Wilson loops along straight lines in ABJM theory or $\mN=4$ CSM quiver theories, different number of Poincar\`e supercharges can be preserved, depending on how scalars and fermions are included in the loop.
It is different from the usual BPS Wilson loops in {\it four-dimensional } $\mN=4$ or $\mN=2$ super Yang-Mills theories, where the BPS Wilson loops along straight lines can only be 1/2 BPS, if we consider only Poincar\'e  supercharges as here.
In $\mN=3$ quiver CSM theories, both DT type and GY type BPS Wilson loops are 1/3 BPS and no further SUSY enhancement appears. This is consistent with the results in the M-theory side \cite{Chen:2014gta}.

Our results inspire quite a few interesting problems for further exploration. It would be nice to figure out the holographic duals of these novel BPS Wilson loops.
It is also worth searching for the origin of these loop operators, beginning with the BEH mechanism as in \cite{Lee:2010hk}. One can also try to construct and study some new BPS cusped Wilson loops \cite{Griguolo:2012iq,Correa:2014aga} using these new DT type BPS Wilson loops as building blocks.
We hope such cusped BPS Wilson loops still play the role as one of the bridges between the localization computations and integrable structure as first proposed in \cite{Correa:2012at}.
It is also quite interesting to investigate whether DT type BPS Wilson loops exist in {\it four-dimensional} $\mN=4$ or $\mN=2$ quiver gauge theories.

\section*{Acknowledgement}
We would like to thank Nan Bai, Bin Chen, Song He, Yu Jia, Wei Li, Jian-Xin Lu, Gary Shiu and Zhao-Long Wang for valuable discussions.
We thank the anonymous referee for suggestions about discussing circular BPS Wilson loops.
The work was in part supported by National Natural Science Foundation of China grants No.~1222549 and No.~11575202. JW also gratefully acknowledges the support of K.~C.~Wong Education Foundation.
JW would also like to thank the participants of the advanced workshop ``Dark Energy and Fundamental Theory'' supported by the Special Fund for Theoretical Physics from National Natural Science Foundation of China with grant No.~11447613 for stimulating discussion.

\appendix

\section{Spinor convention}

We use the same convention as the one in \cite{Ouyang:2015ada}. In three-dimensional Minkowski spacetime we use the coordinates $x^\m=(x^0,x^1,x^2)$, the metric $\eta_{\m\n}=\diag(-++)$, and the gamma matrices
\be
\g^{\m\phantom{\a}\b}_{\phantom{\m}\a}=(\ii\s^2,\s^1,\s^3).
\ee
Charge conjugate of a general spinor is defined as
\be
\bar\th_\a=\th_\a^*,
\ee
with * denoting complex conjugate.

In three-dimensional Euclidean space we use the coordinates $x^\m=(x^1,x^2,x^3)$, the metric $\eta_{\m\n}=\diag(+++)$, and the gamma matrices
\be
\g^{\m\phantom{\a}\b}_{\phantom{\m}\a}=(-\s^2,\s^1,\s^3).
\ee
Generally, spinors $\th$ and $\bar\th$ are not related.


\providecommand{\href}[2]{#2}\begingroup\raggedright\endgroup

\end{document}